\documentclass[referee]{svjour3}
\usepackage{graphicx}
\usepackage{rotating}
\usepackage{amssymb}
\usepackage{mathptmx}
\usepackage[numbers, square, sort]{natbib}
\usepackage{booktabs}
\usepackage{multirow}
\usepackage{floatrow}
\usepackage{caption}
\usepackage{stackengine}
\usepackage{authblk}
\makeatletter
\journalname{}
\begin{document}

\AtEndEnvironment{thebibliography}{
% all your extra bibitems go here
\bibitem{dahal2019} S. Dahal, et al., \textit{Journal of Low Temperature Physics}, This Special Issue (2019).
}

\newcommand{\jcap} {JCAP}
\newcommand{\procspie}{Proc. SPIE} %Proceedings of the International Society for Optical Engineering
\newcommand{\urss}[1]{\ensuremath{_{\mathrm{#1}}}}
\newcommand{\hdblarrow}{H\makebox[0.9ex][l]{$\downdownarrows$}-}

\title{Demonstration of 220/280\,GHz Multichroic Feedhorn-Coupled TES Polarimeter}

\author{S.~Walker$^{1,2}$ \and
C. E.~Sierra$^3$ \and
J. E.~Austermann$^2$ \and
J. A.~Beall$^2$ \and
D. T.~Becker$^{1,2}$ \and
B. J.~Dober$^2$ \and
S. M.~Duff$^2$ \and
G. C.~Hilton$^2$ \and
J.~Hubmayr$^2$ \and
J. L.~Van Lanen$^2$ \and
J. J.~McMahon$^3$ \and
S. M.~Simon$^3$ \and
J. N.~Ullom$^{1,2}$ \and
M. R.~Vissers$^2$
}

\authorrunning{S.~Walker et. al.}

\institute{S.~Walker \\
%2000 Colorado Ave., Boulder, CO 80309 \\
%Tel.: +1-303-497-6483\\
\email{samantha.walker-1@colorado.edu} \\
$^1$ University of Colorado Boulder, Boulder, CO, USA \\
$^2$ National Institute of Standards and Technology, Boulder, CO, USA \\
$^3$ University of Michigan, Ann Arbor, MI, USA \\
}

\maketitle

\begin{abstract}\label{section:abstract}

We describe the design and measurement of feedhorn-coupled, transition-edge sensor (TES) polarimeters with two passbands centered at 220\,GHz and 280\,GHz, intended for observations of the cosmic microwave background.
Each pixel couples polarized light in two linear polarizations by use of a planar orthomode transducer and senses power via four TES bolometers, one for each band in each linear polarization. 
Previous designs of this detector architecture incorporated passbands from 27\,GHz to 220\,GHz; we now demonstrate this technology at frequencies up to 315\,GHz.
Observational passbands are defined with an on-chip diplexer, and Fourier-transform-spectrometer measurements are in excellent agreement with simulations.
We find coupling from feedhorn to TES bolometer using a cryogenic, temperature-controlled thermal source.
We determine the optical efficiency of our device is $\eta = 77\%\pm6\%$ ($75\%\pm5\%$) for 220 (280)\,GHz, relative to the designed passband shapes.
Lastly, we compare two power-termination schemes commonly used in wide-bandwidth millimeter-wave polarimeters and find equal performance in terms of optical efficiency and passband shape.

\keywords{feedhorn, polarimeter, microwave, millimeter-wave, cosmic microwave background, CMB, transition-edge sensor, TES}

\end{abstract}

\section{Introduction}\label{section:intro}

The cosmic microwave background (CMB) provides a powerful probe of the earliest moments of the universe. 
Precision measurements of CMB temperature and polarization aniso-tropies have played a crucial role in shaping our understanding of how the universe formed by providing rigorous constraints \cite{planck2018vi} on parameters of the standard cosmological model, $\Lambda$CDM.
However, millimeter-wave observations are complicated by the presence of astrophysical foregrounds, such as synchrotron emission and galactic dust, which also radiate at these wavelengths. 
By designing detectors with broad spectral coverage, these foregrounds can be separated from the CMB because their spectral energy distributions are distinct. 
In addition, these detectors maximize the usage of telescope focal planes.
For these reasons, several research groups are developing multichroic detectors \cite{mcmahon2012} with different coupling architectures \cite{suzuki2012,anderson2018,dahal2019}. 
We are developing feedhorn-coupled transition-edge sensor (TES) polarimeters, which have been deployed in multiple experiments \cite{ho2017,choi2018,koopman2018} over bands from 27\,GHz to 220\,GHz.
In this work, we describe our first implementation of multichroic, dual-polarization-sensitive 220/280\,GHz detectors for the Simons Observatory~\cite{so2019}.

\begin{figure}[t]
\begin{center}
\includegraphics[width=0.8\linewidth, keepaspectratio]{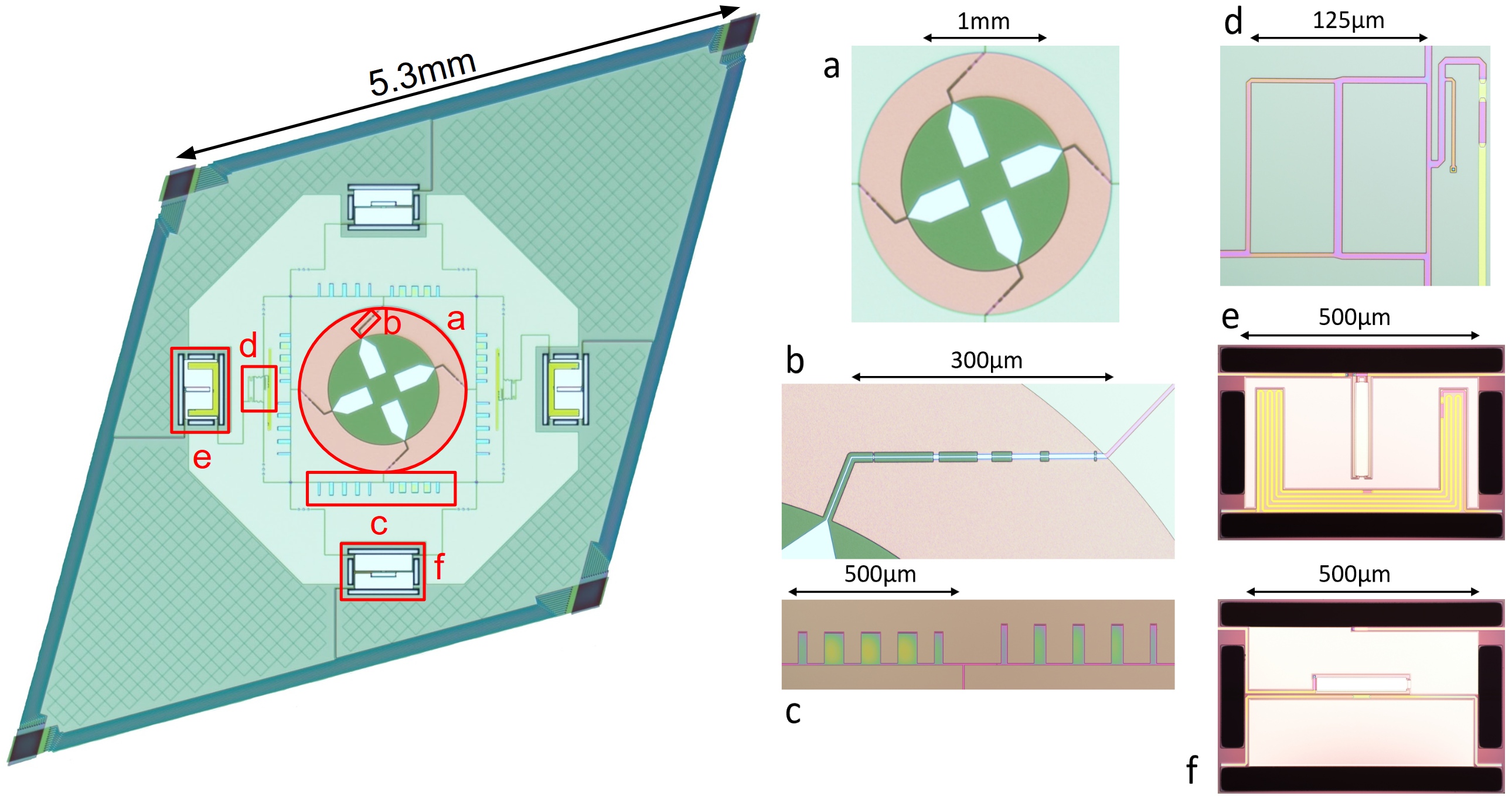}
\caption{{\it Left:} Optical micrograph of 220/280\,GHz multichroic detector with major components denoted by {\it red labels}. {\it Right:} Detailed micrographs of: {\bf a} planar orthomode transducer (OMT), {\bf b} co-planar waveguide-to-microstrip (CPW-to-MS) transition, {\bf c} diplexer, {\bf d} hybrid tee and {\bf e} TES bolometer with Nb-to-Au MS transition for `A-type' termination, and {\bf f} TES bolometer with lumped termination resistor for `B-type' termination. (Color figure online.)}

\label{figure:micrographs}
\end{center}
\end{figure}

\section{Design and Simulated Performance}\label{section:designsim}

We have designed and fabricated single-pixel, feedhorn-coupled, 220/280\,GHz polarimeters. 
Devices were fabricated in the NIST Boulder Microfabrication Facility. 
The general feedhorn-coupled architecture is described by McMahon et al. \cite{mcmahon2012}. 
Fabrication is described in more detail by Duff et al. \cite{duff2016}. 
Here we focus on aspects unique to this design. 
The device consists of four optically-coupled TES bolometers for passband measurements in two linear polarizations.
Additionally, we include two dark TES bolometers for systematic checks. 
TESs consist of an AlMn alloy and were designed to have $T\urss{c}$~=~$160$\,mK and $R\urss{n}$~=~8\,m$\mathrm{\Omega}$.
Detector pixels were also designed with two different bolometer-saturation powers ($P\urss{Sats}$).
The first type has TESs with continuous membranes suitable for room-temperature Fourier-transform-spectrometer (FTS) measurements. 
The second type has TESs with released membranes suitable for ground-based photon loading in the Atacama Desert, Chile.
Fig.~\ref{figure:micrographs} shows an optical micrograph of the detector pixel with components highlighted.

As part of our first implementation, we also designed each polarization with a different power-termination: a hybrid tee coupled to a lossy Au distributed termination (`A-type') \cite{chuss2012} or a PdAu lumped termination resistor (`B-type') \cite{myers2005}. 
The function of this termination is to dissipate only the lowest-order waveguide mode, TE$_{11}$, on the bolometer.
At 2.3:1 bandwidth, the planar orthomode transducer (OMT) supports multiple waveguide modes; however, we only wish to couple the bolometer to the TE$_{11}$ mode because it exhibits a well defined polarization state. 
Furthermore, coupling the bolometer to higher-order modes is undesirable as it degrades angular resolution \cite{hubmayr2018}. 
In the A-type, filtered passbands are routed to a hybrid tee, which takes the input from two ports and produces sum and difference outputs. 
Higher-order modes are associated with the sum output and are terminated on the substrate, while power in the TE$_{11}$ mode associated with the difference port is routed to the bolometer and dissipated in a lossy Nb-to-Au meander.
In the B-type, a lumped termination resistor, thermally connected to the TES, is differentially fed signals from one pair of OMT probes.
This termination scheme also results in a smaller bolometer than in the A-type, which could be beneficial for more tightly packed pixels.

The performance of the components listed in Fig.~\ref{figure:micrographs} was verified through electromagnetic (EM) simulations in Microwave Office (1D, www.awr.com), Sonnet (2.5D, www.-sonnetsoftware.com/), and HFSS (3D, www.ansys.com).
We created a full 3D model in HFSS of the OMT (Fig.~\ref{figure:micrographs}a), including a backshort and waveguide coupling.
We calculated the scattering parameters of five waveports, one for the waveguide input and one for each of the four OMT probes. 
This model assumed no loss so that any power not found in these five waveports was assumed to be due to radiation that could leak out of the 25\,$\mathrm{\mu}$m gaps between the planar OMT above and below the waveguide. 
From this, we predict 95.4\% (89.6\%) co-polar coupling, 2.4\% (2.8\%) reflection, and 2.2\% (7.6\%) radiation, averaged over the 220 (280)\,GHz passband.
The co-planar waveguide-to-microstrip (CPW-to-MS) transition (Fig.~\ref{figure:micrographs}b) was optimized with a transmission-line model in Microwave Office and then verified in Sonnet. 
Mean reflection across the extended passband, 195\,GHz to 315\,GHz, was simulated to be $-25$\,dB. 
The diplexer (Fig.~\ref{figure:micrographs}c) was designed and verified in Sonnet. 
We estimate 94.7\% (91.0\%) transmission assuming SiN loss tangent (tan $\delta$) = 0.0008, averaged over the 220 (280)\,GHz passband.
For the A-type termination, separate hybrid tees (Fig.~\ref{figure:micrographs}d) were designed in Sonnet for the 220\,GHz and 280\,GHz passbands, because they have intrinsically narrow bandwidths. 
From the scattering parameters, the differential-odd mode, associated with the difference port, was calculated to have $>$ 99\% transmission for each type. 
The Nb-to-Au transition (Fig.~\ref{figure:micrographs}e), like the CPW-to-MS transition, was optimized with a transmission-line model in Microwave Office and verified in Sonnet.
Mean reflection across the extended passband was simulated to be $-35$\,dB. 
For the B-type termination, the termination resistor (Fig.~\ref{figure:micrographs}f) was simulated in Sonnet, assuming PdAu with sheet resistance $R = 3.2\,\mathrm{\Omega}$/square and an ideal aspect ratio of length:width $=$ 7:1. 
Mean reflection across the extended passband was simulated to be $-25$\,dB.

\section{Experimental Setup}\label{section:setup}

To test the devices, we packaged the dies in brass split-block modules coupled to prototype aluminum spline-profiled feedhorns \cite{simon2016}, coupled to a time-division multiplexer (TDM) \cite{doriese2016}, and installed in a 100\,mK adiabatic-demagnetization-refrigerator cryostat.
We configured the cryostat in two ways for our two measurements. 
For measurement of detector passbands, we installed an optical-access window in the cryostat and used a Fourier-transform-spectrometer (FTS). 
The optical path from FTS to feedhorn consists of the following filters. 
At 300\,K, we use a 8.89\,cm diameter aperture stop and 5.08\,cm thick polypropylene-based, expanded foam access window. 
At 50\,K, we use a 2\,cm thick single-layer anti-reflection (AR) coated PTFE filter. 
This AR coating has been optimized for transmission from 195\,GHz to 420\,GHz. 
At 4\,K, we use a 14\,cm$^{-1}$ metal-mesh low pass filter \cite{ade2006} from Cardiff, a 1\,cm thick Teflon filter, and a 1.5875\,mm thick nylon stack with the same AR coating as the 50\,K component.
For measurement of detector optical efficiency, we used a cryogenic temperature-controlled thermal source heat-sunk to the 4\,K stage. 
This thermal load consists of a 101.6\,mm tesselating terahertz tile that has $<$~$-30$\,dB reflection at 300\,GHz \cite{saily2004}. 
We therefore assume a perfect blackbody in our loading calculation discussed in the next section. 
Two uncalibrated thermometers on the backside of this thermal load are used to measure temperature changes. The source's exponential decay-time constant was measured to be $\sim2.5$ minutes.
We place 14\,cm$^{-1}$ and 11\,cm$^{-1}$ low pass filters from Cardiff at 4\,K to further attenuate any out-of-band pickup.

\begin{figure}[t] %htbp
\begin{center}
\includegraphics[width=0.8\linewidth, keepaspectratio]{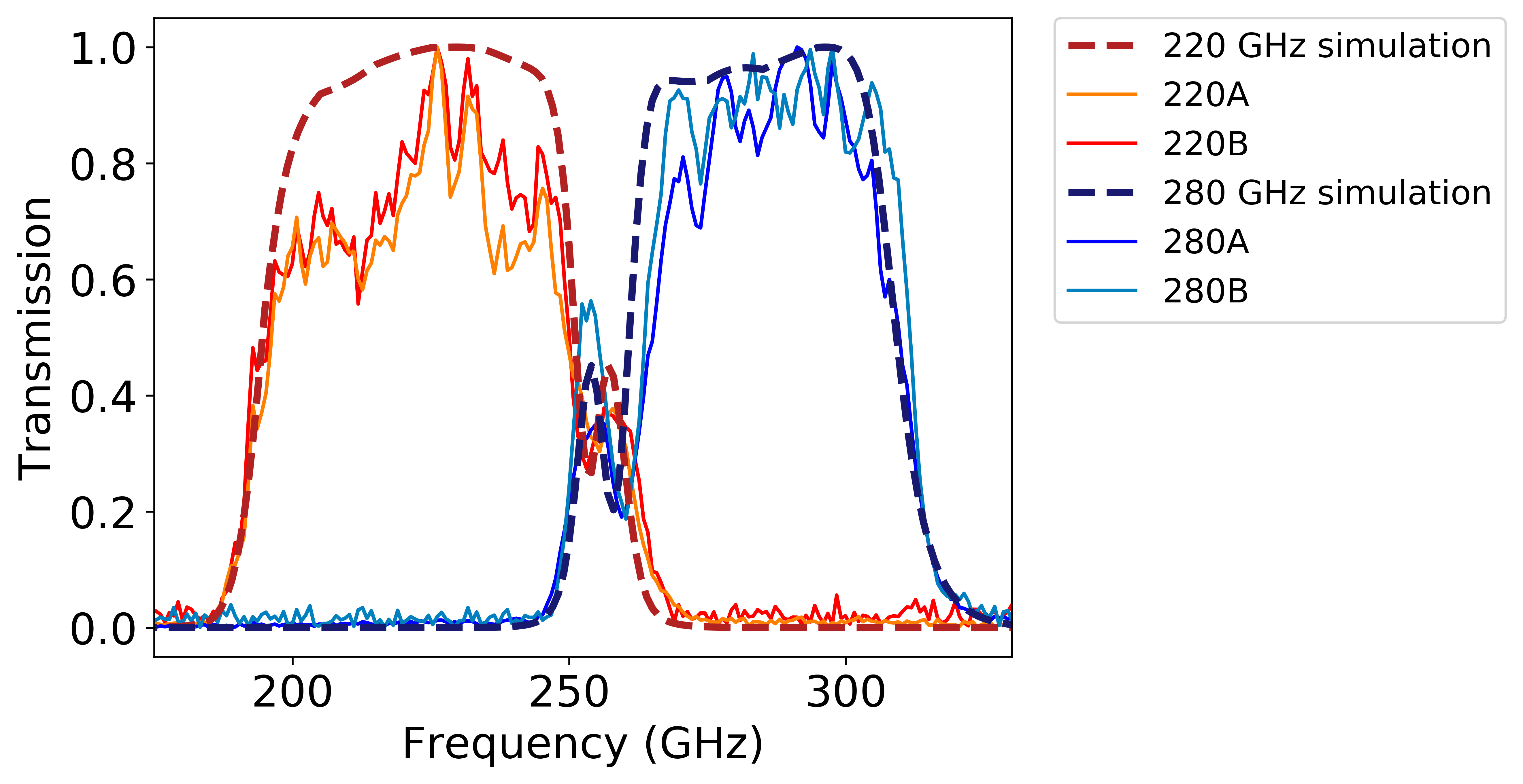}
\caption{Passband measurements of the 220/280\,GHz detector. A and B refer to orthogonal polarizations with different termination schemes described in more detail in Section~\ref{section:designsim}. Data ({\it solid lines}) are in excellent agreement with simulations ({\it dashed lines}). Both data and simulations are peak normalized. (Color figure online.)}

\label{figure:passbands}
\end{center}
\end{figure}

\section{Results}\label{section:results}

\subsection{Passbands}

Fig.~\ref{figure:passbands} shows peak normalized FTS measurements of the 220/280\,GHz detector along with separate 220\,GHz and 280\,GHz simulations that consist of the simulated peak normalized frequency response of the diplexer and OMT. 
There is excellent agreement with simulations and both A- and B-type terminations give well-matched passbands.
This agreement also shows that the slight interaction between the two passbands at $\sim$250\,GHz is understood; this was removed with a simple design change in subsequent wafer production.
Furthermore, the anti-reflection coating used in our filter stack is non-ideal. 
Given our knowledge of the materials and distances in the optical path from FTS to feedhorn, we expect fringing at the level of 20\% to 30\%, which is also seen in the data.

\begin{figure}[t]
\begin{center}
\includegraphics[width=0.6\linewidth, keepaspectratio]{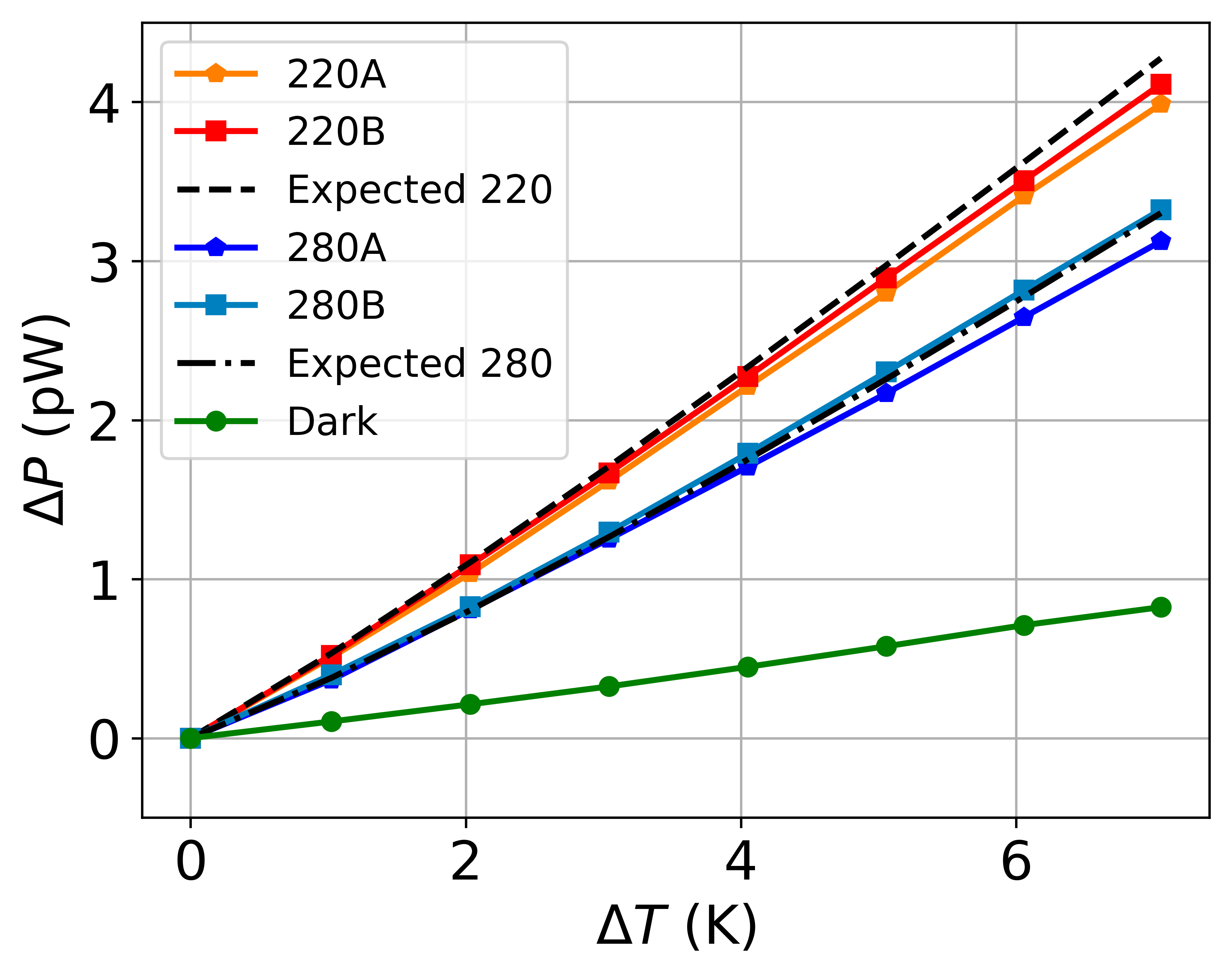}
\caption{Power change in detectors as a function of temperature change of a cryogenic blackbody with reference to 5\,K. {\it Solid lines} denote measurements and {\it dashed lines} correspond to theoretical loading $\Delta P\urss{Load}$. (Color figure online.)}

\label{figure:deltaPvsdeltaT}
\end{center}
\end{figure}

\begin{table}[t]
\begin{center}
		\begin{tabular}{@{} lccr @{}} % Column formatting, @{} suppresses leading/trailing space
			\toprule
			%\multicolumn{2}{c}{Item} \\
			%\cmidrule(r){1-2} % Partial rule. (r) trims the line a little bit on the right; (l) & (lr) also possible
			Type   & Raw & With dark subtraction \\
			\midrule
			220A &  95\%$\pm$1\% & 76\%$\pm$2\% \\
			220B & 97\%$\pm$4\% & 79\%$\pm$6\%  \\
			280A & 97\%$\pm$4\% & 72\%$\pm$4\%  \\
			280B & 103\%$\pm$3\% & 78\%$\pm$4\%  \\
			\bottomrule
		\end{tabular}
\caption{Optical efficiency for raw $\Delta P$ data (`Raw') and after subtracting the contribution to $\Delta P$ from dark TESs (`With dark subtraction'). Error bars are the one standard deviation statistical uncertainty from 42 independent measurements, two pixels, seven $\Delta T$ points, and three repeats, summed in quadrature.}

\label{table:opticalefficiency}
\end{center}
\end{table}

\subsection{Optical Efficiency}

Because CMB measurements are inherently of low signal-to-noise ratio, we seek to maximize the optical efficiency of our detectors. 
\enlargethispage{-\baselineskip}
We define optical efficiency $\eta$ as the ratio of the optical power dissipated in our TES bolometers ($\Delta P$) to the calculated loading power we would expect $\Delta P\urss{Load}$: 

\begin{equation}\label{eq:eta}
	\eta = \Delta P/\Delta P\urss{Load}.
\end{equation}
We vary the temperature of a beam-filling cryogenic blackbody load from 5\,K to 12\,K and measure the change in power in each TES through bolometer $I-V$ curves for $P$ at 80\% $R\urss{n}$. 
Measurements are taken at each 1\,K point after 21 minutes to allow the thermal load to fully equilibrate. 
We calculate $P\urss{Load}$ for a single mode and ideal blackbody, including simulated passbands and corrections for free space filter loss, as

\begin{equation}\label{eq:pload}
	P\urss{Load} = \int \frac{h \nu}{e^{\frac{h\nu}{k\urss{B} T}} - 1}\prod_{i = 1}^{4} F_i(\nu)d\nu,
\end{equation}
where $h$ is the Planck constant, $\nu$ is frequency, $k\urss{B}$ is the Boltzmann constant, and $T$ is the blackbody temperature in kelvin.
$F_i(\nu)$ includes the peak normalized 220\,GHz or 280\,GHz simulated diplexer response, the peak normalized simulated OMT response, and the 300\,K measured absolute transmission spectra of the 14\,cm$^{-1}$ and the 11\,cm$^{-1}$ low pass filters.
Fig.~\ref{figure:deltaPvsdeltaT} shows $\Delta P$ measurements of both optical and dark TES bolometer channels as a function of cryogenic blackbody temperature (referenced to 5\,K) denoted by solid lines. 
Dashed lines correspond to calculations of expected loading power, $\Delta P\urss{Load}$, using Eq.~(\ref{eq:pload}).

Optical efficiencies are calculated with Eq.~(\ref{eq:eta}) for each $\Delta T$ point. 
The average is reported in Table~\ref{table:opticalefficiency}, both with (`With dark subtraction') and without (`Raw') dark bolometer power subtraction. 
Error bars in Table~\ref{table:opticalefficiency} are the one standard deviation statistical uncertainty from 42 independent measurements, two pixels, seven $\Delta T$ points, and three repeats, summed in quadrature. 
Assuming the $\Delta P$ of dark bolometers accurately monitors the parasitic power radiatively coupled to the optically coupled bolometers, we subtract dark $\Delta P$ from optical $\Delta P$ to estimate $\eta$.
We calculate $\eta = 77\%\pm6\%$ ($75\%\pm5\%$) for the 220 (280)\,GHz passband.
Factors that could account for this additional loss include: imprecise knowledge of load temperature, ideal versus measured passband shape, uncertainty in the gap size of the OMT, and the dark subtraction assumption itself.
Given these factors, there is general agreement between measurements and expectations.

\section{Conclusions}\label{section:conclusions}

We have demonstrated a working prototype of a multichroic feedhorn-coupled 220/280\,GHz detector pixel through electromagnetic simulations, FTS passband measurements, and cryogenic temperature-controlled blackbody measurements to calculate optical efficiency. 
Measured passbands are in excellent agreement with simulations. 
We determine detector optical efficiency $\eta = 77\%\pm6\%$ ($75\%\pm5\%$) for the 220 (280)\,GHz passband. 
These results are consistent with expectations and demonstrate efficient optical coupling from feedhorn to TES bolometer. 
As part of our first implementation of these devices, we also investigated two power-termination schemes commonly used in wide-bandwidth millimeter-wave polarimeters: a hybrid tee coupled to a lossy Au distributed termination (A-type) or a PdAu lumped termination resistor (B-type). 
We find equal performance in terms of optical efficiency and passband shape. 

After testing a second prototype of this detector pixel, this design will be implemented in arrays to be fielded in the Simons Observatory.
In arrays to be deployed, an all-silicon assembly will reduce the OMT waveguide gap by more than a factor of 2. 
This will result in both an expected increase in co-polar coupling and a decrease in leakage radiation, improving the optical performance beyond what we demonstrate here. 

\begin{acknowledgements}

This work was supported in part by the NASA APRA program, grant \#NNX17AL23G. 
This material is based upon work supported by the National Science Foundation Graduate Research Fellowship under Grant No. DGE-1144083. 
Certain commercial software and materials are identified to specify the experimental study adequately. 
This does not imply endorsement by NIST nor that the software and materials are the best available for this purpose.

\end{acknowledgements}

\bibliographystyle{unsrt85}
\bibliography{Walker_LTD18_publist}

\end{document}